\documentclass[aps,twocolumn,prc,floatfix,showpacs,preprintnumbers,amsmath,amssymb,nofootinbib,groupedaddress]{revtex4-1}

\usepackage[normalem]{ulem}
\usepackage{wasysym}
\usepackage{color}
\usepackage{graphicx}
\usepackage{dcolumn}    
\usepackage{bigstrut}
\usepackage{amsmath,amsfonts,amsthm,bm}
\usepackage{bm}
\usepackage{CJK}
\usepackage[pdfstartview=FitH,
            CJKbookmarks=true,
            bookmarksnumbered=true,
            bookmarksopen=true,
            colorlinks,
            pdfborder=001,
            linkcolor=blue,
            anchorcolor=blue,
            citecolor=blue
            ]{hyperref}

\def\be{\begin{equation}}
\def\ee{\end{equation}}
\def\bea{\begin{eqnarray}}
\def\eea{\end{eqnarray}}
\def\bse{\begin{subequations}}
\def\ese{\end{subequations}}

\begin{document}
%
\title{Isobaric analog state energy in deformed nuclei: a toy model}

\author{X. Roca-Maza}
\email{xavier.roca.maza@mi.infn.it}
\affiliation{Dipartimento di Fisica ``Aldo Pontremoli'', Universit\`a degli Studi di Milano, 20133 Milano, Italy}
\affiliation{INFN,  Sezione di Milano, 20133 Milano, Italy}

\author{H. Sagawa}
\email{sagawa@ribf.riken.jp}
\affiliation{RIKEN Nishina Center, Wako 351-0198, Japan}
\affiliation{Center for Mathematics and Physics, University of Aizu, Aizu-Wakamatsu, Fukushima 965-8560, Japan}

\author{G. Col\`o}
\email{gianluca.colo@mi.infn.it}
\affiliation{Dipartimento di Fisica ``Aldo Pontremoli'', Universit\`a degli Studi di Milano, 20133 Milano, Italy}
\affiliation{INFN,  Sezione di Milano, 20133 Milano, Italy}

\date{\today} 

\begin{abstract}
A formula to evaluate the effects of a general deformation on the Coulomb direct contribution to the energy of the Isobaric Analog State (IAS) is presented and studied via a simple yet physical model. The toy model gives a reasonable account of microscopic deformed Hartree-Fock-Bogolyubov (HFB) calculations in a test case, and provides a guidance when predicting unknown IAS energies. Thus, deformed HFB calculations, to predict the IAS energies, are performed for several neutron-deficient medium-mass and heavy nuclei which are now planned to be studied experimentally.   
\end{abstract}


\maketitle 

\section{Introduction}
Isospin is one of the most important (approximate) symmetries in nuclei. The validity of isospin symmetry has been established by the experimental observation of isobaric analogue states (IAS) by charge-exchange reactions. Recently,  these states have been investigated extensively in connection with the symmetry energy, in particular to determine the so-called slope parameter $L$ \cite{Dani2017, roca-maza2018a, roca-maza2018b}.  The nuclear symmetry energy is one of the fundamental ingredients to describe the nuclear Equation of State (EoS) when dealing with isospin-asymmetric matter. Its determination may entail profound consequences in our  understanding of various physical observables; the symmetry energy governs not only properties of nuclei, but also  numerous facets of astrophysics like neutron stars and supernovae \cite{BAL,Steiner,Baldo}. The nuclear EoS and symmetry energy have been discussed in many different contexts, in which both the strong and Coulomb interactions play a role.  It should be noticed,  however, that our knowledge of the strong interaction, even in its realistic form employed in ab initio-type calculations, have some room to be improved for reaching a more robust understanding of the nucleus and of the nuclear EoS. On the other hand, the IAS is essentially governed by the well-established Coulomb force, which is an advantage when trying to elucidate the EoS in asymmetric nuclear matter.  
 
Up to now, theoretical studies of IAS have been mainly focused on spherical nuclei such as $^{48}$Ca, $^{90}$Zr and $^{208}$Pb. The only few exceptions, to the best of our knowledge, are the Skyrme calculations by K. Yoshida \cite{Yoshida} and by the Madrid group \cite{alvarez2005,sarriguren2011}. However, a large number of deformed nuclei exist  in wide regions of the mass table, and  play an important role for various nuclear structure problems. While experimental data of IAS exist in some deformed nuclei, the information on deformation effects on the IAS is still missing in a transparent way besides only a few theoretical study The IAS in neutron-rich nuclei have been considered in most of the previous studies since the  large isospin values of these nuclei make it easier to observe experimentally the IAS as an isolated narrow resonance. On the other hand, neutron-deficient nuclei have some advantages since the differences between  the proton and neutron densities (i.e.,  the size of the neutron skin) is relatively small and, consequently, their absolute errors are smaller. This allows evaluating the density differences $\Delta \rho_{np}=\rho_n-\rho_p$ with a better systematic accuracy from charge-exchange reactions.  Another interesting aspect is that   most neutron-deficient isotopes in the medium- to heavy-mass region may have large isospin mixing, which will have a noticiable impact on the IAS energy and its systematics.
  
In this paper, we derive a general formula for the deformation effects on the Coulomb direct contribution to the energy of the IAS and provide a simple albeit physical model. In addition, we estimate the deformation effects using a microscopic Hartree-Fock-Bogolyubov (HFB) model and test both the general formula and the proposed toy model. Then, we study  several  neutron-deficient medium-mass and heavy-nuclei, which are now planned to be studied experimentally in RCNP, Osaka within the LUNESTAR project \cite{LUNESTAR}
  
The paper is organized as follows. In Sec.~\ref{model}, the theoretical model is introduced. In Sec.~\ref{results}, we test the general formula to account for deformation effects on the IAS energy and the toy model by comparing them with HFB results. We also provide and discuss deformed HFB predictions for experimentally accessible neutron-decficient nuclei. Conclusions are drawn in Sec.~\ref{conclusions}.

\section{Model}
\label{model}
\subsection{Definition of the IAS energy}

The Isobaric Analog State (IAS) energy $E_{\rm IAS}$ can be defined as the energy difference between the analog state $\vert A\rangle$ and the parent state $\vert 0\rangle$. The parent state is an eigenstate of the Hamiltonian $\mathcal{H}$ with $N$ neutrons and $Z$ protons and the analog state can be defined as \cite{auerbach1972}
\begin{equation}
  \vert A\rangle\equiv \frac{T_-\vert 0\rangle}{\langle 0\vert T_+T_-\vert 0\rangle^{1/2}}, 
\label{eq1}
\end{equation}
where $T_+=\sum_i^A t_+(i)$ and $T_-=\sum_i^A t_-(i)$ are the isospin raising and lowering operators, respectively, that follow the usual SU(2) algebra:
\begin{equation}
[T_+,T_-]=2T_z ,  \, \, \, \, [T_z,T_{\pm}]=\pm T_{\pm}.
\label{eq2}
\end{equation}
$T_z=\sum_i^A t_z(i)$ and $t_z$ has eigenvalues $-1/2$ for protons and $1/2$ for neutrons. Hence,
\begin{equation}
  E_{\rm IAS} = \langle A\vert \mathcal{H}\vert A\rangle - \langle 0\vert \mathcal{H}\vert 0\rangle = \frac{\langle 0\vert T_+[\mathcal{H},T_-]\vert 0\rangle}{\langle 0\vert T_+T_-\vert 0\rangle} \ .
\label{eq3}
\end{equation}
Assuming good isospin in the parent state $T_+\vert 0\rangle=0$, Eq. \eqref{eq3} is rewritten as 
\begin{equation}
  E_{\rm IAS} = \frac{1}{N-Z}\langle 0\vert [T_+,[\mathcal{H},T_-]]\vert 0\rangle \ .
\label{eq4}
\end{equation}
It is important to note that the latter formula can be only applied to nuclei with $N>Z$. If isospin mixing is considered, namely if $T_+\vert 0\rangle\ne 0$, Eq.~(\ref{eq4}) should be corrected as in Eq.~(A6) of Ref.~\cite{roca-maza2020}. A simple approximate expression for that correction is \cite{auerbach1972} (cf. also Appendix A in Ref.~\cite{roca-maza2020})  
\begin{equation}
E_{\rm IAS}^{\rm IM}\approx -170\varepsilon^2\frac{N-Z+2}{N-Z}A^{-1/3}, 
\label{eq5}
\end{equation}
where $\varepsilon$ is the isospin mixing in the parent state. Specifically, if the parent state wave function is $\vert 0\rangle = \sum_n a_n\vert T_0+n, T_0\rangle$, and the states $\vert T_0+n, T_0\rangle$ have good isospin, then $\varepsilon\equiv a_1$: the admixture of states with total isospin $T>T_0+1$ is expected to be very small and can be neglected. Under this assumption, we can define
\begin{equation}
\varepsilon^2 \equiv \frac{\langle 0\vert T_-T_+\vert 0\rangle}{N-Z+2}
\label{epsilon0}
\end{equation}
where, for our purposes here, it is accurate to evaluate the numerator on a Bardeen-Cooper-Schrieffer (BCS) ground state wave function and assuming spherical symmetry\footnote{It is important to note here that deformation would be a correction to isospin mixing which is a correction to the IAS energy. The approximate character of Eq.~(\ref{eq5}) is consistent with the fact that we are dealing with a second-order, small effect.}. This will give the simple and closed expression,  
\begin{equation}
\varepsilon^2 \equiv \frac{1}{N-Z+2}\sum_{\substack{n_p,n_n\\l_p=l_n\\j_p=j_n}}(2j_p+1)v_p^2u_n^2\mathcal{O}_{np}^2
\label{epsilon}
\end{equation}
with the overlap factor between the neutron ($n$) and proton ($p$) radial part $R_{n,l}(r)$ of the considered single particle wave function 
\begin{equation}
  \mathcal{O}_{np} \equiv \int_0^\infty dr r^2 R_{n_p,l_p}(r) R_{n_n,l_n}(r) \ .
\label{gamma}
\end{equation}
In these expressions, $n$ is the principal quantum number, and $l$ and $j$ are the orbital and total angular momentum quantum numbers, respectively, while $v$ and $u$ corresponds to the occupation factors. As a test, we have confirmed that the estimation in Eq.~(\ref{epsilon0}) calculated within the Tamm-Dancoff approximation coincides within a very good accuracy (1\% or below) with the expression in Eq.~(\ref{epsilon}).

\subsection{Contributions to the IAS energy}

Due to the structure of Eq. (\ref{eq3}), $E_{\rm IAS}$ depends only on isospin symmetry breaking (ISB) 
parts of $\mathcal{H}$. In nuclear physics, the main isospin breaking term is known to be due 
to the Coulomb interaction. Therefore, the bulk contribution to Eq. (\ref{eq3}) will be due to the 
difference in the expectation value of the Coulomb matrix elements between proton and neutron distributions. That is, for the direct Coulomb (Cd) term, assuming an independent particle model, one has
\begin{equation}
E_{\rm IAS}^{\rm Cd} = \frac{1}{N-Z}\int \left[\rho_n({\bm r})-\rho_p({\bm r})\right]U_{\rm C}^{\rm direct}({\bm r}) d{\bm r}, 
\label{eq6}
\end{equation}
where $U_{\rm C}^{\rm direct}({\bm r})$ is the direct part of the Coulomb  potential generated by the electric charge distribution $\rho_{\rm ch}({\bm r})$,
\begin{equation}
U_{\rm C}^{\rm direct}({\bm r}) = \int\frac{e^2}{\vert{\bm r}^\prime-{\bm r}\vert}\rho_{\rm ch}({\bm r}^\prime) d{\bm r}^\prime. 
\label{eq7}
\end{equation}
Under the same assumption and adopting, in addition, the Local Density Approximation (LDA), the Coulomb exchange contribution can be also conveniently written as a function of the neutron and proton density distributions as   
\begin{equation}
E_{\rm IAS}^{\rm Cx} = \frac{1}{N-Z}\int \left[\rho_n({\bm r})-\rho_p({\bm r})\right]U_{\rm C}^{\rm exch, LDA}({\bm r}) d{\bm r}, 
\label{eq8}
\end{equation}
where $U_{\rm C}^{\rm exch, LDA}({\bm r})$ is the Coulomb exchange part of the Coulomb energy potential generated by the electric charge distribution $\rho_{\rm ch}({\bm r})$ within the LDA,
\begin{equation}
U_{\rm C}^{\rm exch, LDA}({\bm r}) = -e^2\left[\frac{3}{\pi}\rho_{\rm ch}({\bm r})\right]^{1/3} \ .
\label{eq9}
\end{equation}
This contribution will be much smaller than the Coulomb direct part. In what follows we will approximate $\rho_{\rm ch}({\bm r})$ by $\rho_{\rm p}({\bm r})$. This approximation produces a negligible 
effect for our purposes here, but should be dropped if one wishes a precise value of the IAS energy \cite{roca-maza2018a,roca-maza2018b}. QED corrections in the fine-structure constant to the Coulomb potential and Coulomb correlation effects will be also neglected in the present study.  

Other ISB effects than Coulomb force come from the nuclear strong interaction. Those terms can be parametrized by simple effective interactions solved at the Hartree-Fock level as it has been recently shown \cite{roca-maza2018a,roca-maza2018b,baczyk2018,baczyk2019}. For example, assuming a Charge Symmetry Breaking (CSB) and Charge Independence Breaking (CIB) effective interaction written as
\begin{eqnarray}
V_{\rm CSB}({\bm r}_1, {\bm r}_2) &=& \frac{1}{2}[t_z(1)+t_z(2)]s_0(1+y_0P_\sigma) \delta({\bm r}_1 - {\bm r}_2), 
\label{eq10}\nonumber\\
V_{\rm CIB}({\bm r}_1, {\bm r}_2) &=& 2t_z(1)t_z(2)u_0(1+z_0P_\sigma) \delta({\bm r}_1 - {\bm r}_2)\ ,
\label{eq11}
\end{eqnarray}
in analogy to the well known Skyrme interaction and where $P_\sigma$ is the exchange operator in spin-space, one finds the following contributions to Eq.~(\ref{eq4}):  
\begin{eqnarray}
E_{\rm IAS}^{\rm CSB} &=& -\frac{1}{4}\frac{s_0(1-y_0)}{N-Z}\int d{\bm r}\left[\rho_n^2({\bm r})-\rho_p^2({\bm r})\right],  
\label{eq12}\\
E_{\rm IAS}^{\rm CIB} &=& -\frac{1}{2}\frac{u_0(1-z_0)}{N-Z}\int d{\bm r}\left[\rho_n({\bm r})-\rho_p({\bm r})\right]^2. 
\label{eq13}
\end{eqnarray}
Hence, the total IAS energy can be accurately estimated by taking into acount all these contributions \cite{roca-maza2018a,roca-maza2018b}. In the present study, we will focus only on the terms that are commonly included in current EDFs, that is, Coulomb direct and Coulomb exchange. In addition, since we will base our microscopic calculations on a Hartree-Fock-Bogoliubov (HFB) approach, we should correct for the spurious isospin mixing. To do that we will resort to the approximate formula given in Eq.~(\ref{eq5}).

Finally, the IAS energy will be estimated from Eqs.~(\ref{eq5}), (\ref{eq6}) and (\ref{eq8}) as
\begin{equation}
E_{\rm IAS}=E_{\rm IAS}^{\rm Cd}+E_{\rm IAS}^{\rm Cx}+E_{\rm IAS}^{\rm IM} \ .
\label{eq14}
\end{equation}

\subsection{Deformation effects to the IAS energy}

In the following we explicitly evaluate Eqs.~(\ref{eq6}) and (\ref{eq8}) assuming the neutron and proton densities, $\rho_n({\bm r})$ and $\rho_p({\bm r})$, can be deformed and relating this result to the spherical case. To this end, we will follow the approach given in Ref.~\cite{carlson1961}.

We start by introducing some notation and writing the square of the distance vector ${\bm R}$ of a deformed system (ellipsoid) as $R^2=(x/a)^2+(y/b)^2+(z/c)^2$ where $x$, $y$ and $z$ are Cartesian coordinates and $a$, $b$ and $c$ dimensionless quantities such that $abc=1$ and, thus, the length of each semi-axis is $aR$, $bR$ and $cR$. The spherical case is recovered for $a=b=c=1$. The relation between the modulus of the deformed distance vector $R$ and the modulus of the spherical distance vector $r$ is
\begin{eqnarray}
  R^2&=&r^2\left(\frac{\sin^2\theta\cos^2\phi}{a^2}+\frac{\sin^2\theta\sin^2\phi}{b^2}+\frac{\cos^2\theta}{c^2}\right)\nonumber \\
  &\equiv& r^2S(\theta,\phi).     
\label{eq15}
\end{eqnarray}  
We are considering a general family of deformations that conserve the volume. $d{\bm R} = dRd\Omega = drd\Omega=d{\bm r}$ since the Jacobian of the transformation is 1. Then, $\int d^3r \rho^\alpha(R) = \int d^3R \rho^\alpha(R) = \int d^3r \rho^\alpha(r)$ or, in other words, integrals that depend on an arbitrary power $\alpha$ of the density will not depend on deformation, as long as one does not introduce an explicit dependence on $r$. Therefore, the result of Eqs.~(\ref{eq8}) --as well as those in Eqs.~(\ref{eq12}) and (\ref{eq13})-- is independent of deformation effects. Only the Coulomb direct contribution shown in Eq.~(\ref{eq6}) will display some dependence on deformation.

To deal with the direct Coulomb contribution, we will compare the result of Eq.~(\ref{eq6}) between a spherical nucleus with densities $\rho_n(r)$ and $\rho_p(r)$ and a deformed nucleus assuming that the neutron and proton densities satisfy $a_n\approx a_p$, $b_n \approx b_p$ and $c_n \approx c_p$ and have the same functional form of the corresponding spherical nucleus but depending on $R_n^2\equiv r^2S_n(\theta,\phi)$ and $R_p^2\equiv r^2S_p(\theta,\phi)$. That is, we shall write $\rho_n(R_n)$ and $\rho_p(R_p)$.

In order to evaluate Eq.~(\ref{eq6}) for a deformed nucleus within the conditions above described, we will first perform a Fourier transform
\begin{eqnarray}
  E_{\rm IAS}^{\rm Cd} &=& \frac{e^2}{N-Z}\frac{1}{2\pi^2}\int d{\bm R}_nd{\bm R^\prime}_p\nonumber\\&&\int d{\bm q} \left[\rho_n(R_n)-\rho_p(R_p)\right]\frac{e^{\imath {\bm Q}_n{\bm R}_n}e^{-\imath {\bm Q}_p{\bm R}_p^\prime}}{q^2} \rho_p(R_p^\prime) \ .\nonumber\\
\label{eq16}
\end{eqnarray}  
By defining ${\bm Q}\equiv(aq_x,bq_y,cq_z)$ in analogy with ${\bm R}\equiv(x/a,y/b,z/c)$, that is, $Q_n^2=q^2/S_n(\theta,\phi)$ and $Q_p^2=q^2/S_p(\theta,\phi)$ and by taking into account that $d{\bm R}_n=d{\bm r}$ and $d{\bm R}_p=d{\bm r}$, we can write
\begin{eqnarray}
&&  E_{\rm IAS}^{\rm Cd} = 8\frac{e^2}{N-Z}\int\frac{d{\bm q}}{q^2} \int dR_p^\prime {R_p^\prime}^2\rho_p(R_p^\prime) j_0(Q_pR_p^\prime)\Big\{\nonumber\\
  && \int dR_n R_n^2\rho_n(R_n)j_0(Q_nR_n) - \int dR_p R_p^2\rho_p(R_p)j_0(Q_pR_p)\Big\} \ . \nonumber\\
\label{eq17}
\end{eqnarray}  
Since $Q_nR_n=Q_pR_p$, the last expression can be written as
\begin{eqnarray}
E_{\rm IAS}^{\rm Cd} &=& 8\frac{e^2}{N-Z}\int\frac{d{\bm q}}{q^2} \int dR_p^\prime {R_p^\prime}^2\rho_p(R_p^\prime) j_0(Q_pR_p^\prime)\nonumber\\
  &&~~~~~~ \int dR_p R_p^2\left[\lambda^3\rho_n(\lambda R_p)-\rho_p(R_p)\right]j_0(Q_pR_p) \ , \nonumber\\
\label{eq18}
\end{eqnarray}  
where
\begin{equation}
  \lambda\equiv \frac{Q_p}{Q_n}=\left[\frac{S_n(\theta,\phi)}{S_p(\theta,\phi)}\right]^{1/2}
\label{eq19}
\end{equation}
depends only on the angles, and $\lambda^3\rho_n(\lambda R)$ corresponds to a volume conserving scaling of the neutron density $\rho_n(R)$. For $\lambda=1$ we obtain the result for equally deformed neutron and proton density distributions. Eq.~(\ref{eq18}) can be further developed as
\begin{eqnarray}
E_{\rm IAS}^{\rm Cd} &=& 8\frac{e^2}{N-Z}\int \frac{d\Omega}{S_p(\theta,\phi)}\int dQ_p dR_p^\prime {R_p^\prime}^2\rho_p(R_p^\prime) j_0(Q_pR_p^\prime)\nonumber\\
  &&~~~~~~ \int dR_p R_p^2\left[\lambda^3\rho_n(\lambda R_p)-\rho_p(R_p)\right]j_0(Q_pR_p) \ . \nonumber\\
\label{eq20}
\end{eqnarray}  
We now perform the following manipulations: i) change of variables $\tilde{R}_p=\lambda R_p$ in the integral that goes with $\rho_n$,
\begin{equation}
\int d\tilde{R}_p \tilde{R}_p^2\rho_n(\tilde{R}_p)j_0(\frac{Q_p}{\lambda}\tilde{R_p}) ; \nonumber  
\end{equation}  
ii) expand $j_0$ for $\lambda \rightarrow 1$ assuming similar neutron and proton deformations -- as previously stated -- and keep the lowest order in $\lambda$, 
\begin{eqnarray}
  j_0(\frac{Q_p}{\lambda}\tilde{R_p})&\approx& j_0(Q_p\tilde{R_p})+\left[j_0(Q_p\tilde{R_p})-\cos(Q_p\tilde{R_p})\right](\lambda-1)\nonumber\\
  &\approx& \lambda j_0(Q_p\tilde{R_p})-(\lambda-1)\cos(Q_p\tilde{R_p}) ; \nonumber
\end{eqnarray}  
iii) perform the change of variables$R_p=\tilde{R}_p$,  
\begin{eqnarray}
&&\lambda \int dR_p  R_p^2\rho_n(R_p)j_0(Q_pR_p)\nonumber\\  
&&-(\lambda-1) \int dR_p  R_p^2\rho_n(R_p)\cos(Q_pR_p).\nonumber  
\end{eqnarray}  
By substituting the last expression in Eq.~(\ref{eq20}), we find
\begin{eqnarray}
  E_{\rm IAS}^{\rm Cd} &=& E_{\rm IAS}^{\rm Cd,sph} \frac{1}{4\pi}\int d\Omega\frac{\lambda}{S_p(\theta,\phi)} \nonumber\\
  &+&8\frac{e^2}{N-Z}\int d\Omega\frac{1-\lambda}{S_p(\theta,\phi)}\int dQ_p\Big\{\nonumber\\
  &&\int dR_p^\prime {R_p^\prime}^2\rho_p(R_p^\prime) j_0(Q_pR_p^\prime)\nonumber\\
  &&\int dR_p R_p^2\left[\rho_n(R_p)\cos(Q_pR_p)-\rho_p(R_p)j_0(Q_pR_p)\right]\Big\}, \nonumber\\
\label{eq21}
\end{eqnarray}  
where $E_{\rm IAS}^{\rm Cd, sph}$ corresponds to the result assuming spherically symmetric neutron and proton densities. Neglecting the term in $\lambda-1$ that should be close to zero under our assumptions, we finally obtain
\begin{eqnarray}
  E_{\rm IAS}^{\rm Cd} &=& E_{\rm IAS}^{\rm Cd,sph}\frac{1}{4\pi}\int d\Omega\frac{\lambda}{S_p(\theta,\phi)}  \nonumber\\
  E_{\rm IAS}^{\rm Cd} &=& E_{\rm IAS}^{\rm Cd,sph}\frac{1}{4\pi}\int d\Omega\frac{[S_n(\theta,\phi)]^{1/2}}{[S_p(\theta,\phi)]^{3/2}}  \ . 
\label{eq22}
\end{eqnarray}
The last expression differs from the result obtained assuming spherical symmetry of the neutron and proton density distributions by a factor
\begin{equation}
\frac{1}{4\pi}\int d\Omega \frac{[S_n(\theta,\phi)]^{1/2}}{[S_p(\theta,\phi)]^{3/2}} \ .
\label{eq23}  
\end{equation}
Assuming equally deformed neutron and proton distributions ($\lambda=1$) the factor would be
\begin{equation}
  \frac{1}{4\pi}\int \frac{d\Omega}{S(\theta,\phi)}, 
\label{eq24}  
\end{equation}
where we have dropped the subindex for obvious reasons. 

These expressions are valid for {\it any} deformation that preserves the volume, and where the neutron and proton distributions are deformed in a similar way. To a good approximation, this is the case for most deformed nuclei according to available microscopic calculations \cite{gogny}. Hence, the usefulness of Eq.~(\ref{eq22}).

In order to give some example in terms of the parameters commonly used in nuclear physics, one should better write $S(\theta,\phi)$ in terms of spherical harmonics,
\begin{equation}
S(\theta,\phi)^{-1/2} = \sum_{lm} \alpha_{lm}Y_{lm}(\theta,\phi) + \mathcal{C}, 
\label{eq25}
\end{equation}  
taking care of renormalizing the expression so that the volume is preserved by determining the proper value of the constant $\mathcal{C}$. That is, by imposing
\begin{eqnarray}
  \int d{\bm r} \rho(r) &=&\int d{\bm R} \rho(R)\nonumber\\
  1 &=& \frac{1}{4\pi}\int \frac{d\Omega}{S(\theta,\phi)^{3/2}}. \nonumber\\  
\label{eq26}
\end{eqnarray}  
As an example, if we assume quadrupole deformation defining
\begin{equation}
\left[S(\theta,\phi)\right]^{-1/2} = 1 + \beta_2Y_{20} + \mathcal{C}, 
\label{eq27}
\end{equation}  
where $\alpha_{20}\equiv \beta_2$, the value of $\mathcal{C}$ is now $-\beta_2^2/4\pi$. Hence, 
it is easy to evaluate the effect on the energy of the IAS as
\begin{eqnarray}
  E_{\rm IAS}^{\rm Cd} &=& E_{\rm IAS}^{\rm Cd,sph} \left[1-\frac{\beta_{2n}\beta_{2p}}{4\pi}\right.\nonumber\\
    &&~~~~~~ +\left.\frac{(\beta_{2n}-\beta_{2p})(\beta_{2n}+\beta_{2p})}{4\pi}+\frac{(\beta_{2n}-\beta_{2p})^2}{4\pi}\right] \ . \nonumber\\
\label{eq28}
\end{eqnarray}  
For the special case $\beta_{2n}=\beta_{2p}$ this reduces to 
\begin{equation}
E_{\rm IAS}^{\rm Cd}=E_{\rm IAS}^{\rm Cd, sph}\left[1-\frac{\beta^2_2}{4\pi}\right], 
\label{eq29}
\end{equation}
where we have neglected the terms in $\beta_2^4$. From the result in Eq.~(\ref{eq28}), one should expect that {\it the larger the quadrupole deformation the smaller the IAS energy}. In order to have a qualitative idea about the effect of deformation on the IAS energy, Eq.~(\ref{eq29}) predicts a realtive difference that goes as $-\beta_2^2/4\pi$, which means, for very deformed nuclei with $\beta_2\approx 0.8$, a relative reduction of $E_{\rm IAS}$ of about 5\% with respect to the spherically symmetric case. So, in general, {\it deformation effects to the IAS energy are expected to be small, about few \% at most}.  

Since it will be useful for testing purposes in the next section, let us use Eq.~(\ref{eq29}) to define an effective quadrupole deformation assuming as known the Coulomb direct energies of an axially deformed nucleus and its spherical counterpart,
\begin{equation}
\beta_2^{\rm eff}\equiv 4\pi\left(1-\frac{E_{\rm IAS}^{\rm Cd}}{E_{\rm IAS}^{\rm Cd, sph}}\right)^{1/2} \ .
\label{eq29a}
\end{equation}

\subsection{Toy model}

In order to understand in simple terms the relation between the $E_{\rm IAS}$, the nuclear quadrupole deformation and the neutron skin thickness of a spherical system $\Delta r^{\rm sph}_{\rm np}\equiv \langle r_n^2\rangle^{1/2}_{\rm sph}-\langle r_p^2\rangle^{1/2}_{\rm sph}$, we evaluate $E_{\rm IAS}^{\rm Cd, sph}$ as in Refs.~\cite{roca-maza2018a,roca-maza2018b} within a simple yet physical model. Within such model that assumes a uniform neutron and proton spherical distributions (sharp sphere approximation) \cite{roca-maza2018a}, one finds  
\begin{eqnarray}
  E_{\rm IAS}^{\rm Cd}&\approx& \left(1-\frac{\beta^2_2}{4\pi}\right) E_{\rm IAS}^{\rm Cd,sph}\nonumber\\
  &\approx&\frac{6}{5}\sqrt{\frac{3}{5}}\frac{Ze^2(1-\beta^2_2/4\pi)}{\langle r_p^2\rangle^{1/2}_{\rm sph}}\left(1-\frac{1}{2}\frac{N}{N-Z}\frac{\Delta r^{\rm sph}_{\rm np}}{\langle r_p^2\rangle^{1/2}_{\rm sph}}
\right), \nonumber\\
\label{eq30}
\end{eqnarray}
that is, the IAS energy should decrease with both increasing neutron skin thickness and nuclear quadrupole deformation. For simplicity, within our toy model, we will assume $\beta_{2n}=\beta_{2p}$. 

It is interesting to note that the effect of quadrupole deformation on the mean square radius is   
\begin{eqnarray}
  \langle r^2\rangle &=&\int d{\bm r} r^2 \rho(R)\nonumber\\
  &=&\int d{\bm r} r^2\rho\left(r\left[1+\beta_2Y_{20}-\frac{\beta_2^2}{4\pi}\right]^{-1}\right)\nonumber\\
  &=&(1+\frac{5}{4\pi}\beta_2^2)\langle r^2\rangle_{\rm sph} \ .
\label{eq31}
\end{eqnarray}
From the latter result, it is clear that the effect of deformation on the energy of the IAS cannot be accounted for only by taking into account the deformation effects on the rms radii of protons and neutrons, although this will produde the largest effect. By using the result in Eq.~(\ref{eq31}) we can rewrite Eq.~(\ref{eq30}), to order $\beta_2^2$, as follows
\begin{equation}
  E_{\rm IAS}^{\rm Cd}\approx\frac{6}{5}\sqrt{\frac{3}{5}}\frac{Ze^2(1+\frac{3}{8\pi}\beta^2_2)}{\langle r_p^2\rangle^{1/2}}\left(1-\frac{1}{2}\frac{N}{N-Z}\frac{\Delta r_{np}}{\langle r_p^2\rangle^{1/2}}\right) \ . \\
\label{eq32}
\end{equation}
This expression allows one to directly use experimentally known rms radii and deformations to estimate $E_{\rm IAS}^{\rm Cd}$. It should be noticed that the deformation increases the numerator, while the charge radius in the denominator is increased even more by the deformation than the numerator.  The net effect of deformation decreases the IAS energy \eqref{eq32}, consistently with Eq. \eqref{eq30}.  

Within this model we can also derive a simple formula for the Coulomb exchange contribution of Eq.~(\ref{eq8}). The result reads as follows, 
\begin{eqnarray}
  E_{\rm IAS}^{\rm Cx}&=&-\sqrt{\frac{3}{5}}\left(\frac{3}{2\pi}\right)^{2/3}\frac{e^2Z^{1/3}}{\langle r_p^2\rangle^{1/2}}\left(1-3\frac{N}{N-Z}\frac{\Delta r_{np}}{\langle r_p^2\rangle^{1/2}}\right) \ .\nonumber\\
  \label{eq33}
\end{eqnarray}

\section{Results}
\label{results}

\subsection{Deformation effects on a test nucleus}

\begin{figure}[h!]
\vspace{5mm}  
\includegraphics[width=\linewidth,clip=true]{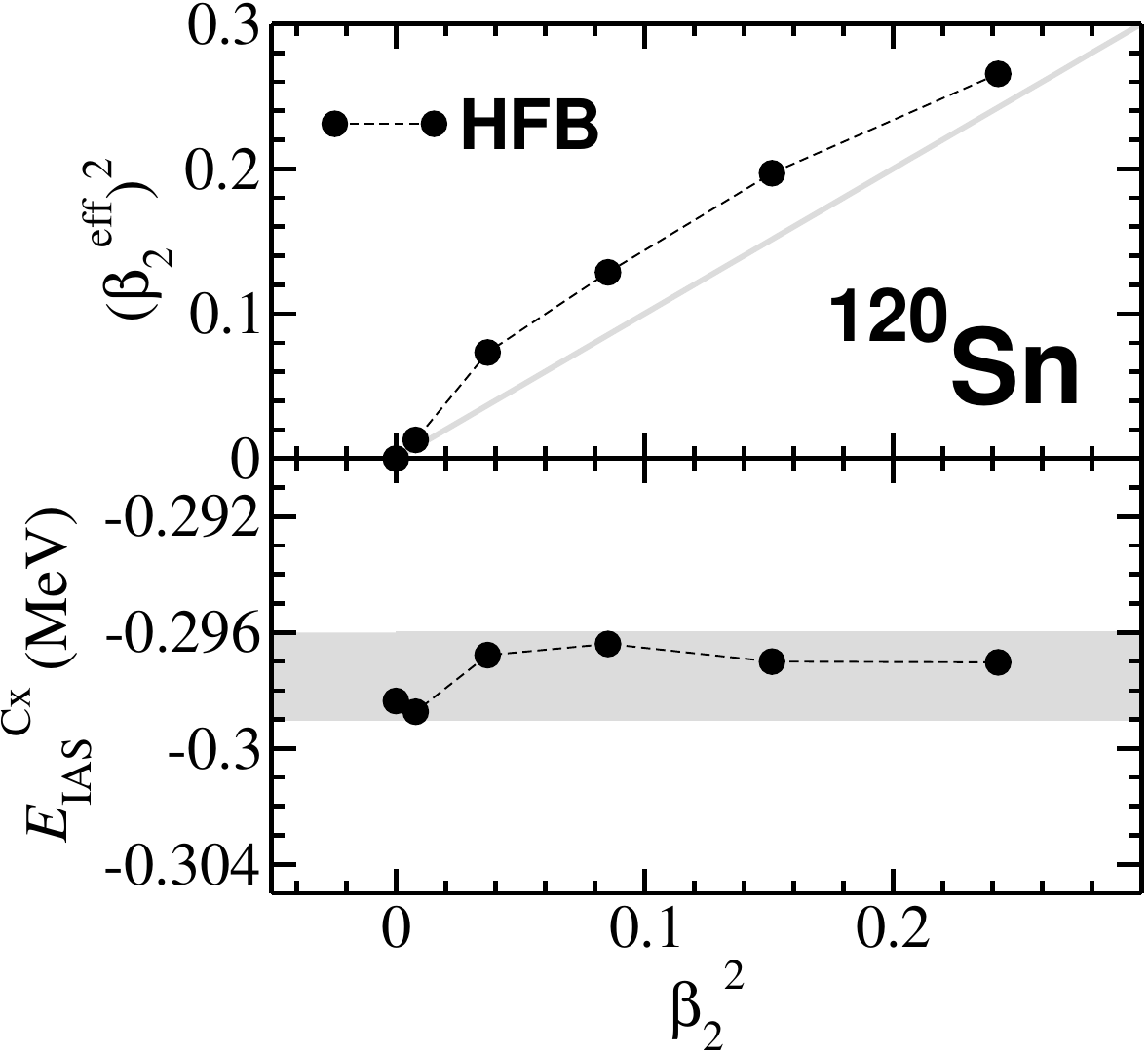}
\caption{Upper panel: $\beta_2^{\rm eff}$ in Eq. \eqref{eq29a} as a function of the mass deformation $\beta_{2m}$. The gray line corresponds to the limit $\beta_2^{\rm eff}=\beta_{2m}$. Lower panel: $E_{\rm IAS}^{\rm Cx}$ as a function of $\beta_{2m}$. The gray zone is just for a guide to eyes. All calculations have been performed with the SAMi functional.}
\label{fig1}
\end{figure}

In this subsection, we have performed different axially deformed constrained HFB calculations for the ${}^{120}$Sn nucleus, from $\beta_2=0$ to $\beta_2=0.5$. We have slightly modified the code HFBTHO \cite{Stoitsov} to use the SAMi interaction and calculate the displacement energies shown in this paper. The number of oscillator shells of the basis is 20, and we calculate pairing correlations with particle number projection after variation, with a density dependent pairing interaction of the type
\begin{equation}
V(\textbf{r}_{1},\textbf{r}_{2}) = V_{0}\left[1 + x
\left(
\frac{\rho(\frac{\textbf{r}_{1}+\textbf{r}_{2}}{2})}{\rho_{0}}
\right)
\right]\delta(\textbf{r}_{1}-\textbf{r}_{2}),
\label{vpair}
\end{equation}
with $x=0.5$ (mixed pairing). The strength $V_0$ is fixed as $V_0 = -432.5$ MeV fm$^3$ in order to reproduce the neutron pairing gap in $^{120}$Sn when the pairing cut-off energy is 60 MeV.

In Fig.~\ref{fig1}, the upper panel, we show $(\beta_2^{\rm eff})^2$ from Eq. (\ref{eq29a}) as a function of $\beta_{2m}^2$, where 
\begin{equation}
\beta_{2q} = \sqrt{\frac{\pi}{5}}\frac{Q_{2q}}{\langle r_q^2\rangle N_q}, 
\end{equation}
with $q=n, p, m$, 
and 
\begin{equation}
Q_{2m}=Q_{2n}+Q_{2p}
\end{equation}
from the HFB calculations. Note that $\beta_{2m}^2$ is obtained from the HFB densities and could be different from the parameter defining the deformation of the HFB potential. This is a consistency test of Eq.~(\ref{eq29}), or a consistency test between densities and Coulomb potentials. Results in the upper panel of Fig.~\ref{fig1} do not deviate substantially from Eq.~(\ref{eq29}), i.e. from the grey solid line. This result suggests that our model evaluation of deformation effects on the IAS is quite acceptable. Note that we have assumed $\beta_{2n}=\beta_{2p}$ to define $\beta_2^{\rm eff}$ while HFB calculations give some difference for yhe axial quadrupole deformat of neutrons and protons. 

In the lower panel of the same figure, we check numerically the contribution to the IAS energy from the Coulomb exchange term in LDA approximation. The grey area delimits the numerical variation of the Coulomb exchange (within Slater approximation) with deformation. As we discussed in the previous subsection, in principle this variation should be zero. Numerically we find an error of few keV which is also quite acceptable.

\subsection{Toy model test: the Sn isotopic chain}

\begin{figure}[t!]
\vspace{5mm}  
\includegraphics[width=\linewidth,clip=true]{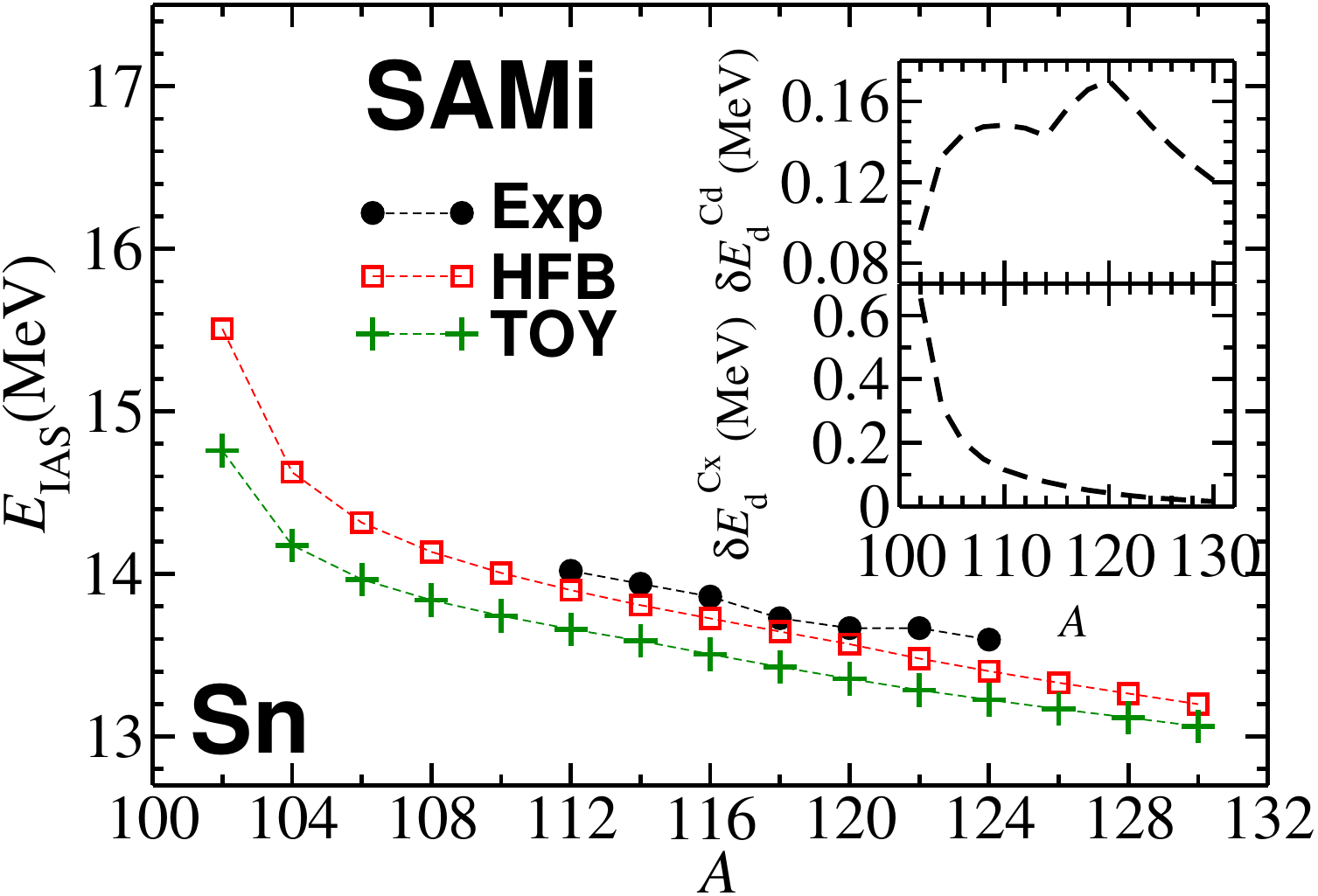}
\caption{Comparison of the HFB results (empty squares) with the toy model results (plus symbols) as predicted by the SAMi interaction along the Sn isotopic chain. Experimental data are also shown when available \cite{antony1997}. In the insets, the absolute deviation of the Coulomb direct term between HFB (\ref{eq6}) and the toy model (\ref{eq32}) are shown in the upper panel, and those between the Coulomb exchange of HFB ~(\ref{eq8}) and the toy model (\ref{eq33}) are shown in the lower panel.}   
\label{fig2}
\end{figure}

In Fig.~\ref{fig2}, we compare the HFB results for the Coulomb direct and exchange contributions in  Eqs.~(\ref{eq6}) and (\ref{eq8}), respectively, with their corresponding toy model counterparts, that is, the direct and exchange ones in Eqs.~(\ref{eq32}) and (\ref{eq33}), respectively.

In Table \ref{tab0} we first show the isospin mixing probabilities, both without and with pairing correlations. Those have been calculated as in Eq.~(\ref{epsilon}), from the overlap of the ground state single-particle neutron and proton wave functions. The laters have been evaluated within the HF-BCS approach, using the same type of pairing interaction and model space as in the HFB calculations so that to reproduce the neutron pairing gap in ${}^{120}$Sn. This approximation is not expected to significatively change the numerical value of $\varepsilon$ and, thus, it is enough for our purposes here to estimate the isospin mixing in the wave function. In connection to that, it is also important to remind the approximate character of our estimate of the isospin mixing contribution to the energy of the IAS by means of  Eq.~(\ref{eq5}). Hence, the values in Table \ref{tab0} should be taken as semi-quantitative results.  More precise results removing spurious mixings by using the Quasi-particle Random Phase Approximation --that exactly restore spurious contributions-- will be discussed in a future publication. As it is expected, the isospin mixing probabilities are larger in neutron-deficient nuclei and smaller in neutron-rich nuclei. As previously investigated \cite{moya2003, alvarez2005}, pairing may enhance the effect of isospin impurites as it is actually found in our case as well. The effect of the isospin mixing on the IAS energy amounts to $-$1.7 MeV in $^{102}$Sn at the largest, and $-$150 keV in $^{120}$Sn at the smallest.

The differences between the microscopic HFB calculations and the macroscopic toy model are shown in the insets of Fig. \ref{fig2}. Our toy model is quite reasonable in the description of the Coulomb  direct part (with an error of 1\% at most), and it reproduces the correct trend of the IAS energy, while the estimate of the Coulomb exchange term is satisfactory only for stable and neutron-rich nuclei. The reason might be twofold: sharp sphere approximation in the toy model; and the larger relevance of surface effects in the Coulomb exchange term when compared to the Coulomb direct one. The latter can be seen from Eqs.~(\ref{eq32}) and (\ref{eq33}) where the {\it surface} correction predicted by the Toy model --term that goes with the neutron skin-- is 6 times larger for the Coulomb exchange.

In summary, the toy model deviation from the self-consistent HFB calculations is about 5\% for ${}^{102}$Sn and decreases smoothly up to 1\% for ${}^{130}$Sn as it an be seen from Fig.~\ref{fig2}.

\begin{table}[t!]
\caption{Isospin mixing probability $\varepsilon^2$ in the Sn isotopes. The energy $E^{\rm IM}_{\rm IAS}$ is calculated by using \eqref{eq5}, with the mixing probability $\varepsilon^2$ in the case with pairing included. Energies are in MeV. Some calculations without pairing did not reach convergence, and correspond to the entries in the Table that are left blank}
\label{tab0}
\centering
\begin{tabular}{ccccccc}
\hline\hline
  $A$& $\varepsilon^2$  (\%)& $\varepsilon^2$ (\%)&$E^{\rm IM}_{\rm IAS}$ (MeV)\\
   &  (w/o pairing)  & (with pairing)  &     \\
   \hline
102 &2.161 &2.363 &$-$1.719\\
104 &1.513 &1.576 &$-$0.854\\
106 &1.238 &1.148 &$-$0.549\\      
108 &      &0.886 &$-$0.395\\
110 &0.700 &0.715 &$-$0.304\\
112 &0.593 &0.597 &$-$0.245\\
114 &0.546 &0.512 &$-$0.205\\
116 &0.436 &0.453 &$-$0.177\\
118 &      &0.418 &$-$0.160\\
120 &0.393 &0.403 &$-$0.152\\
122 &0.392 &0.400 &$-$0.149\\
124 &0.395 &0.402 &$-$0.148\\
126 &0.402 &0.407 &$-$0.148\\
128 &0.411 &0.412 &$-$0.148\\
130 &0.421 &0.420 &$-$0.150\\
\hline\hline
\end{tabular}
\end{table}

\begin{table*}[t!]
\caption{Results of deformed HFB calculations with the SAMi EDF. All energies are in MeV and proton rms radii and neutron skin thickness in fm. $E_{\rm IAS}^{\rm HFB}$ is the sum of the direct Coulomb, exchange Coulomb and the isospin mixing contributions in Eqs.~(\ref{eq5}), (\ref{eq6}) and (\ref{eq8}). The deformation effect on the IAS energy has been estimated in different ways: from the HFB calculations [$\Delta E_{\rm IAS}^{\rm HFB} \equiv E_{\rm IAS}^{\rm HFB}(\beta_{2n},\beta_{2p})-E_{\rm IAS}^{\rm HFB}(\beta_{2n}=0,\beta_{2p}=0)$]; by using HFB calculations to input the quantities appearing in the r.h.s. of Eq. (\ref{eq28}); and by using HFB calculations to input the quantities appearing in the r.h.s. of Eq. (\ref{eq29}) (for the Coulomb direct term in both cases).}
\label{tab1}
\centering
\begin{tabular}{rcccccccrrr}
\hline\hline
Nucl.& $\beta_{2n}$&$\beta_{2p}$&$\beta_{2m}$&$\langle r_p^2\rangle^{1/2}$&$\Delta r_{np}$& $E_{\rm IAS}^{\rm HFB}$& $E_{\rm IAS}^{\rm exp}$ &$\Delta E_{\rm IAS}^{\rm HFB}$&$\Delta E_{\rm IAS}$ & $\Delta E_{\rm IAS}$ \\
 & & & & & & & Ref.~\cite{antony1997} & & [Eq.(\ref{eq28})] &  [Eq.(\ref{eq29})]\\
\hline  
\hline  
${}_{46}^{102}$Pd& 0.186 & 0.174 & 0.180 &4.421 &0.042 &13.061& 	       &$-$0.162 &$-$0.054 &$-$0.036 \\
${}_{48}^{106}$Cd& 0.255 & 0.256 & 0.256\footnote{Deformed secondary energy minimum for $\beta_2\sim 0.15$ at $\Delta E\sim 1$ MeV} &4.514 &0.033 &13.561&	       &$-$0.089 &$-$0.072 &$-$0.074 \\
${}_{50}^{112}$Sn& 0.192 & 0.199 & 0.195\footnote{Spherical secondary energy minimum at $\Delta E\sim 1.5$ MeV} &4.563 &0.046 &13.905&14.019(20)&$-$0.001 &$-$0.032 &$-$0.044 \\
${}_{52}^{120}$Te& 0.039 & 0.045 & 0.042 &4.620 &0.074 &14.237&	       &$-$0.001 &$-$0.005 &$-$0.002 \\
${}_{54}^{124}$Xe& 0.000 & 0.000 & 0.000\footnote{Flat energy surface up to $\beta_{2m}\sim 0.2$ with $\Delta E\sim 1.5$ MeV} &4.677 &0.063 &14.697&	       &0.000 &0.000 &0.000 \\
${}_{56}^{130}$Ba& 0.000 & 0.000 & 0.000 &4.744 &0.069 &14.954&	       &0.000 &0.000 &0.000 \\
${}_{58}^{136}$Ce& 0.126 & 0.158 & 0.140\footnote{Spherical secondary energy minimum at $\Delta E\sim 1$ MeV} &4.819 &0.075 &15.057&	       &$-$0.081 &$-$0.026 &$-$0.024 \\
${}_{58}^{138}$Ce& 0.040 & 0.047 & 0.043 &4.820 &0.087 &15.067&	       &$-$0.010 &$-$0.006 &$-$0.002 \\
${}_{60}^{142}$Nd& 0.000 & 0.000 & 0.000 &4.864 &0.079 &15.540&	       &0.000 &0.000 &0.000 \\
${}_{62}^{144}$Sm& 0.000 & 0.000 & 0.000 &4.893 &0.061 &16.008&16.075(15)&0.000 &0.000 &0.000  \\
${}_{66}^{156}$Dy& 0.201 & 0.225 & 0.211 &5.066 &0.072 &16.499&	       &$-$0.003 &$-$0.016 &$-$0.061 \\
${}_{66}^{158}$Dy& 0.197 & 0.217 & 0.205 &5.076 &0.087 &16.444&	       &$-$0.028 &$-$0.020 &$-$0.057  \\
${}_{68}^{162}$Er& 0.348 & 0.378 & 0.360\footnote{Deformed secondary energy minimum for $\beta_{2m}\sim 0.2$ at $\Delta E\sim 1$ MeV} &5.207 &0.076 &16.743&16.861(16)&$-$0.112 &$-$0.111 &$-$0.181   \\
${}_{68}^{164}$Er& 0.346 & 0.377 & 0.359\footnote{Deformed secondary energy minimum for $\beta_{2m}\sim 0.2$ at $\Delta E\sim 1$ MeV} &5.220 &0.089 &16.668&16.778(9) &$-$0.110 &$-$0.107 &$-$0.178	\\
${}_{70}^{168}$Yb& 0.385 & 0.413 & 0.396 &5.293 &0.086 &16.860&	       &$-$0.212 &$-$0.156 &$-$0.223   \\
${}_{72}^{174}$Hf& 0.304 & 0.319 & 0.310 &5.289 &0.091 &17.426&          &$-$0.091 &$-$0.103 &$-$0.138	\\
${}_{72}^{176}$Hf& 0.267 & 0.276 & 0.271 &5.281 &0.104 &17.341&17.388(7) &$-$0.086 &$-$0.087 &$-$0.105	\\
${}_{74}^{180}$W & 0.278 & 0.299 & 0.286 &5.334 &0.093 &17.718&	       &$-$0.101 &$-$0.073 &$-$0.120  \\
${}_{76}^{184}$Os& 0.335 & 0.355 & 0.343 &5.418 &0.085 &18.035&	       &$-$0.171 &$-$0.126 &$-$0.176  \\
${}_{78}^{190}$Pt&-0.147 &-0.141 &-0.145 &5.359 &0.106 &18.468& 	       &$-$0.037 &$-$0.025 &$-$0.032 \\
${}_{80}^{196}$Hg&-0.180 &-0.187 &-0.183 &5.437 &0.105 &18.736& 	       &$-$0.070 &$-$0.058 &$-$0.051   \\
\hline\hline
\end{tabular}
\end{table*}

\subsection{Neutron-deficient medium- and heavy-nuclei and IAS energies}

In Table \ref{tab1}, we show the results of deformed HFB calculations  as predicted by SAMi for several neutron-deficient nuclei which are now planned to be studied experimentally by LUNESTAR project \cite{LUNESTAR}. $E_{\rm IAS}^{\rm HFB}$ is the sum of the direct Coulomb, exchange Coulomb and the isospin mixing contributions in Eqs.~(\ref{eq5}), (\ref{eq6}) and (\ref{eq8}). The deformation effect on the IAS energy has been estimated in different ways: from the HFB calculations $\Delta E_{\rm IAS}^{\rm HFB} \equiv E_{\rm IAS}^{\rm HFB}(\beta_{2n},\beta_{2p})-E_{\rm IAS}^{\rm HFB}(\beta_{2n}=0,\beta_{2p}=0)$; by using HFB calculations to input the quantities appearing in the r.h.s. of Eq.~(\ref{eq28}); and by using HFB calculations to input the quantities appearing in the r.h.s. of Eq.~(\ref{eq29}). The deformation effect varies from $-$212 keV in $^{168}$Yb at the maximum to zero when the nucleus is not deformed like, for example, in the case of $^{142}$Nd. In general, the deformation effect is small, nevertheless it might be important --together with other neglected contributions not object of this study-- for a precise prediction of the IAS energy. 

The IAS of these nuclei are planned to be measured by charge-exchange $(^3$He$,t)$ reaction at $E_{lab}$ = 420 MeV at RCNP, Osaka University, under the LUNESTAR project \cite{LUNESTAR}. This projectile energy is very selective to excite isospin states with respect to spin-isospin states. The experimental campaign will give us new information of IAS states outside of the valley of stability. Specifically, it is planned to investigate the available most neutron deficient stable nuclei, and the $(^3$He,$t)$ reaction will lead to unstable daughter nuclei. Using $(^3$He,$t)$ ensures the best spectral resolution and the most precise determination of the IAS excitation energy in the daughter nucleus. The mass of the daughter nucleus will be measured by a Penning trap experiment. The two experiments, $(^3$He,$t)$ and the mass measurement,  will be combined to extract the excitation energy of IAS with high accuracy. The results shown in Table \ref{tab1} reproduce  the few experimentally known IAS energies within an error of about 100 keV and, thus, may give a good guide for experimental search of new IAS states of these neutron-deficient nuclei.

The nuclei listed in Table \ref{tab1} have relatively small neutron skin size, so that accurate experimental cross section measurements will not only provide the neutron skin size (small) but also the radial dependence of skin density $\Delta \rho_{np} = \rho_{n}- \rho_{p}$.  Then, the deformation effect may play an important role to establish the link between $\Delta \rho_{np}$ and the symmetry energy in these unstable nuclei, while the contributions to the absolute IAS energy are rather small.  

\section{Summary}
\label{conclusions}
We have studied the isobaric analog state in spherical and deformed nuclei in the medium- and heavy-mass region. We propose a general formula [Eq.~(\ref{eq22})] to account in a simple way for the effects of deformation on the energy of the IAS and a toy model based on such formula to explore both, deformation effects on IAS energy and its dependence with the neutron skin thickness. We examine the validity of the presented model by comparing it with deformed HFB calculations in Sec.~\ref{results}. We have found that the expression (\ref{eq22}) describe well deformation effects and that the toy model works well to describe the Coulomb direct contribution --that is, the main contribution-- to the IAS energy, within an accuracy at the 1\% level. The toy model expression for the Coulomb exchange within the Slater approximation gives also a good account in the stable Sn isotopes, but it shows non-negligible differences for Sn isotopes with $N\sim Z$. 

We have also performed deformed HFB calculations of many neutron-deficient nuclei outside the valley of stability. Our model reproduces well the empirical IAS energies, for those nuclei whose IAS energies have been already measured. Thus, our HFB results may provide a reasonable guide for future experiments in the neutron-deficient nuclei proposed in the LUNESTAR project at RCNP, Osaka University \cite{LUNESTAR}.
  
\begin{acknowledgments}
We would like to thank A. Tamii for encouraging us to study the deformed IAS energies. We thank also K. Yoshida for illustrative discussions on charge exchange QPA for IAS states.  Thanks are for Dieter Frekers, and Tommi Eronen for informing us details of experimental project LUNESTAR.  This work was supported in part by JSPS KAKENHI  Grant Numbers JP19K03858. Funding from the European Union's Horizon 2020 research and innovation programme under grant agreement No 654002 is also acknowledged.
\end{acknowledgments}

\bibliography{bibliography.bib}

\begin{thebibliography}{19}
\bibitem [{LUN()}]{LUNESTAR}%
  \BibitemOpen
  \href@noop {} {}\bibinfo {note} {LUNESTAR. A collaboration project has been
  formed to measure IAS in neutron-deficient nuclei by combination ($^3$He,$t$)
  reactions and high precision mass spectrometry}\BibitemShut {NoStop}%
\end{thebibliography}

\end{document}